\documentclass[journal]{IEEEtran}
\usepackage{amsmath,amsfonts}
\usepackage{algorithmic}
\usepackage{algorithm}
\usepackage{array}
\usepackage[caption=false,font=normalsize,labelfont=sf,textfont=sf]{subfig}
\usepackage{textcomp}
\usepackage{stfloats}
\usepackage{url}
\usepackage{verbatim}
\usepackage{graphicx}
\usepackage{cite}
\usepackage{caption}
\usepackage{multirow}

\begin{document}

\title{Invisible Backdoor Attack Through Singular Value Decomposition}
%
%
%

\author{Wenmin Chen, Xiaowei Xu
\thanks{Wenmin Chen is with Ocean University of China,Qingdao,China
(email: cwm@ouc.edu.cn).}
}

\maketitle

\begin{abstract}
With the widespread application of deep learning across various domains, concerns about its security have grown significantly. Among these, backdoor attacks pose a serious security threat to deep neural networks (DNNs). In recent years, backdoor attacks on neural networks have become increasingly sophisticated, aiming to compromise the security and trustworthiness of models by implanting hidden, unauthorized functionalities or triggers, leading to misleading predictions or behaviors. To make triggers less perceptible and imperceptible, various invisible backdoor attacks have been proposed. However, most of them only consider invisibility in the spatial domain, making it easy for recent defense methods to detect the generated toxic images.To address these challenges, this paper proposes an invisible backdoor attack called DEBA. DEBA leverages the mathematical properties of Singular Value Decomposition (SVD) to embed imperceptible backdoors into models during the training phase, thereby causing them to exhibit predefined malicious behavior under specific trigger conditions. Specifically, we first perform SVD on images, and then replace the minor features of trigger images with those of clean images, using them as triggers to ensure the effectiveness of the attack. As minor features are scattered throughout the entire image, the major features of clean images are preserved, making poisoned images visually indistinguishable from clean ones. Extensive experimental evaluations demonstrate that DEBA is highly effective, maintaining high perceptual quality and a high attack success rate for poisoned images. Furthermore, we assess the performance of DEBA under existing defense measures, showing that it is robust and capable of significantly evading and resisting the effects of these defense measures.
\end{abstract}

%
\IEEEpeerreviewmaketitle

\section{INTRODUCTION}
%
%
%
%
\IEEEPARstart{D}{eep} Neural Networks (DNNs) have demonstrated their efficacy and find extensive applications in pivotal domains like image classification\cite{he2016deep}, image segmentation\cite{feng2022fiba}, and object recognition\cite{peng2021conformer}. Presently, achieving a well-performing model necessitates substantial data and computational resources. Consequently, practitioners often resort to leveraging third-party training resources, such as open-source datasets, cloud computing platforms, and pre-trained models, to alleviate the training burden. However, the convenience of these resources may introduce novel security vulnerabilities. Recent research has underscored the susceptibility of DNNs to backdoor attacks\cite{gu2019badnets}, wherein adversaries surreptitiously embed hidden triggers or modify model parameters during the training phase. This implantation enables poisoned models to behave normatively on clean data but make malicious predictions on tainted samples. Consequently, backdoor attacks pose a grave security menace to DNNs, particularly in safety-critical domains like autonomous driving\cite{nguyen2021wanet}, medical imaging analysis\cite{feng2022fiba}, and military applications\cite{drager2023backdoor}.

 The ideal backdoor attack should function normally when processing benign data, be able to recognize predefined triggers, and execute the attacker's intended actions, while the triggers appear natural or indistinguishable to the naked eye to evade manual inspection. Additionally, it should possess some robustness to evade existing defenses, as the model performs relatively poorly on benign data, making it likely for victims to detect the attack.
 
\begin{figure}[!t]
\centering
\includegraphics[width=3.5in]{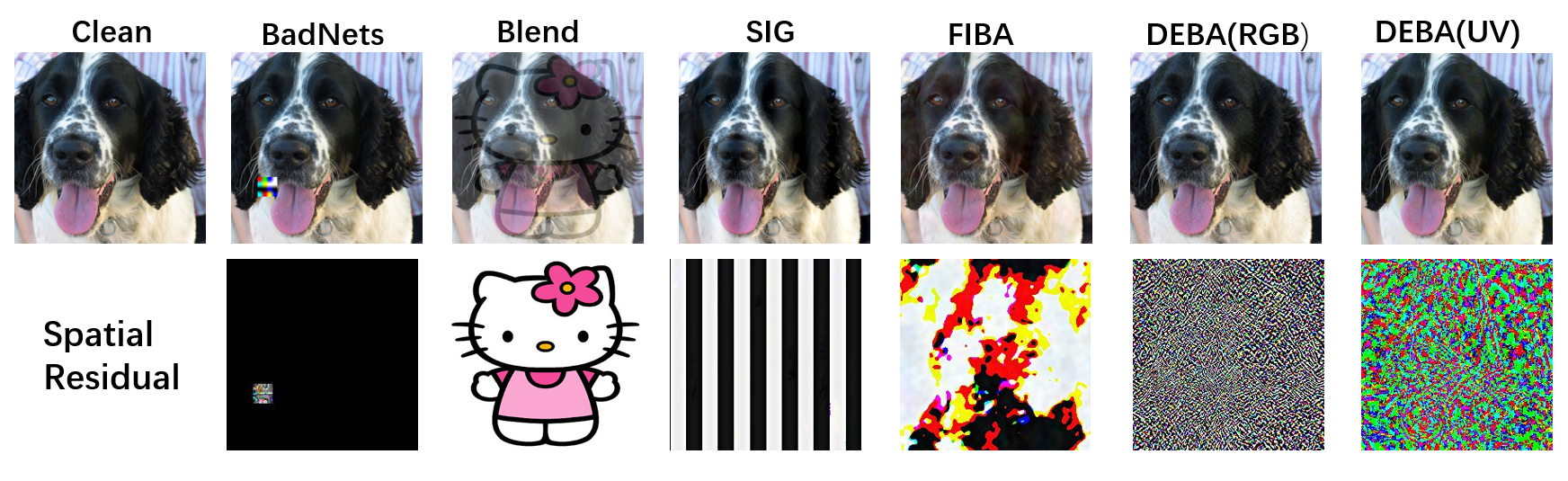}
\caption{
Visualization of different attack methods. The first row of the upper panel shows, from left to right: clean image, toxic images generated by BadNets\cite{gu2019badnets}, Blend\cite{chen2017targeted}, SIG\cite{barni2019new}, and FIBA\cite{feng2022fiba}, as well as two toxic images generated by our proposed DEBA method. The second row displays their corresponding spatial domain residual images, namely the trigger images in the spatial domain.}
\label{Fig.1}
\end{figure}

Existing backdoor attacks mostly involve the fabrication and embedding of triggers in the spatial domain. As depicted in Fig.\ref{Fig.1}, using visible backdoor triggers such as (BadNets\cite{gu2019badnets} and Blend\cite{chen2017targeted}) can achieve the attack objectives, but they can be easily detected and filtered out by manual inspection. Moreover, if some simple data transformations (such as contracted flipping and padding) are applied to preprocess the trigger images, most visible backdoor attacks will completely fail during the inference stage, which is difficult to achieve in practice. In order to evade manual inspection and enhance attack concealment, some propose using methods like distortion fields\cite{nguyen2021wanet} and physical reflectionto\cite{liu2020reflection} make the triggers invisible. However, compared to the original image, there is still a sense of anomaly, and it is not completely indistinguishable to the naked eye.

To address the aforementioned issues, we propose DEBA, a novel invisible backdoor attack based on singular value decomposition (SVD)\cite{kalman1996singularly}. It avoids directly creating backdoor triggers in the spatial domain, making poisoned images almost indistinguishable from clean images to the naked eye, significantly enhancing the stealthiness of the attack. Specifically, we perform singular value decomposition on each channel of both trigger images and clean images (images can be converted to channels other than RGB, such as YUV). Each set of singular values and singular vectors contains different image information, with minor image features corresponding to low singular values and their corresponding singular vectors. We embed the lower singular values and corresponding singular vectors of the trigger images into clean images to obtain poisoned images. Because singular value information can be captured by neural networks, the SVD-based backdoor attack becomes feasible. Moreover, the poisoned images retain a significant amount of the main information of clean images, making the differences between poisoned and clean images almost non-existent and indistinguishable to the naked eye. To validate the effectiveness and reliability of our attack method, we conduct extensive and extensive experiments on benchmark datasets and DNNs. The experimental results show that DEBA has a high attack success rate (ASR) while ensuring a high clean data accuracy (CDA). Additionally, it can effectively evade existing defense methods.

The main contributions of this paper are summarized as follows:
 
\begin{itemize}
\item[$\bullet$] We propose an invisible backdoor attack called DEBA, which injects the backdoor information through singular value decomposition, thus avoiding backdoor embedding in the pixel space domain and achieving ideal invisibility.
\item[$\bullet$] We employ different forms of color channels for backdoor injection.
\item[$\bullet$] Extensive experiments demonstrate that DEBA outperforms several common backdoor attacks in terms of higher attack success rates without compromising the original accuracy, while also surpassing them in robustness and concealment.
\end{itemize}

\section{RELATED WORK}
\subsection{Backdoor Attacks}

Poisoning attacks were first proposed by Xiao et al.\cite{xiao2012adversarial}, who utilized label flipping to induce misclassification in Support Vector Machine (SVM) models. Subsequently, Huang et al.\cite{xiao2015support} further refined attacks by modifying data labels, affecting both linear and nonlinear SVMs, thereby drawing widespread attention to the security issues of SVMs. Although these attacks can lead to inference errors, the low validation accuracy on clean data may hinder the practical deployment of such models.

Recently, the success of backdoor attacks has once again brought widespread attention to the security issues of neural networks (DNNs). Initially proposed by Gu et al.\cite{gu2019badnets}, BadNets embeds a small patch into clean images to form toxic images, and then replaces some of the training data with toxic images to train the model. This causes the trained model to make directional errors during the prediction phase, thus controlling the behavior of the model. Building upon this, Chen et al.\cite{chen2017targeted} proposed a hybrid approach, blending trigger backdoor images and benign images in certain proportions. Additionally, they confirmed that slight noise does not affect the triggering of the backdoor, further reducing the detection risk. However, due to significant differences between toxic and original images, it is challenging to evade manual inspection. Subsequently, there have been a series of efforts dedicated to investigating stealthy backdoor attacks. Turner et al.\cite{turner2019label} introduced an efficient yet label-consistent backdoor attack method, utilizing adversarial perturbations and generative models to generate toxic images.Shafahi et al.\cite{shafahi2018poison}inspired by watermarking, proposed a clean-label attack strategy. Their method creates seemingly plausible but difficult-to-classify inputs, making the model heavily rely on the trigger. Following this, in \cite{li2020invisible}, an adversarial attack was proposed, creating covert and scattered triggers for invisible backdoor creation through steganography and regularization in trigger generation.Later, in order to make the trigger appear more natural, Zhang et al.\cite{zhang2022poison}proposed the "Poison Ink" method, using the inherent edge information as dynamic triggers. Additionally, Jiang et al.\cite{jiang2023color} introduced a new color backdoor attack, demonstrating both robustness and stealthiness. Their key insight involves applying a unified color space shift to all pixels as triggers. This global feature is robust to image transformation operations, and triggered samples maintain a natural appearance. However, upon closer observation, it is apparent that the toxic images generated by these attacks still differ from clean original images and can be visually identified, thus undermining their invisibility. Furthermore, some methods focus on revealing the potential threats of backdoor attacks in the physical world \cite{xue2022ptb}\cite{li2021backdoor}. Zhao et al.\cite{zhao2022natural} conducted backdoor attacks by simulating the effect of raindrops in the real world, while Emily et al.\cite{wenger2020backdoor} constructed a polluted face recognition dataset using physical objects such as glasses and stickers for attack purposes.
\subsection{Backdoor Defenses}
With the rapid development of stealthy backdoor attacks, researchers are continuously seeking effective methods to safeguard neural networks from the impact of such attacks. Consequently, research in backdoor defense has become increasingly active. Presently, backdoor defense mainly focuses on defense against data input, defense against models, and detection of model outputs.

The defensive emphasis is placed on detecting anomalies in inputs. Grad-CAM\cite{selvaraju2017grad}uses saliency maps to analyze the regions of input images that the model focuses on to determine if the image is poisoned. Zeng et al. proposed FTD \cite{zeng2021rethinking}, which uses Discrete Cosine Transform (DCT) to differentiate whether input images have high-frequency artifacts, thereby detecting poisoned images.

In terms of model defense, Liu et al. proposed Fine-Pruning\cite{liu2018fine}, which involves fine-tuning and pruning the model to eliminate backdoors. Wang et al. introduced a method called Neural Cleanse\cite{wang2019neural}, which employs multiple mitigation techniques such as input filtering, neuron pruning, and forget-me-not recognition to identify and reconstruct triggers, and then suppresses the influence of reconstructed triggers to eliminate backdoors. Building upon this, Zhao et al.\cite{zhao2020bridging} proposed using model connectivity techniques to eliminate hidden backdoors in infected models. Li et al. proposed NAD\cite{li2021neural}, while Yoshida et al.\cite{yoshida2020disabling}suggested using model distillation techniques to cleanse backdoor models. Later, Zheng et al.\cite{zheng2022data} introduced a novel defense method called CLP, which can detect potential backdoor channels in a data-free manner and perform simple pruning on infected DNNs to repair them. Building on NAD, Xia et al. proposed ARGD\cite{xia2022eliminating}, an effective backdoor defense algorithm based on graph knowledge distillation, which can eliminate more backdoor triggers than NAD.

Model output detection mainly focuses on detecting anomalies in the outputs. Huang et al. \cite{huang2020one} proposed the effectiveness of single-pixel signatures in detecting classification backdoors in CNNs. Gao et al. proposed STRIP \cite{gao2019strip}, which overlays various image patterns and observes the randomness of predicted classes of perturbed inputs from a given deployed model to test the model, where higher poisoning probability results in lower output randomness. Backdoor attacks and defenses can be seen as two mutually stimulating opposing forces. In this paper, we compare our method and demonstrate that our attack is able to withstand backdoor defenses.

\section{METHOD}

\subsection{Threat Model}

\subsubsection{Attacker’s Capacities}
Current backdoor attacks can target various stages of model deployment, with the training phase being the most common and susceptible to attacks. In this paper, we consider attackers having access to and tampering with a portion of the dataset, but not controlling the training process or the architecture of the target model. That is, the training phase cannot access or modify any parameters of the model. During the inference phase, attackers can only manipulate input images to test the trained model. This threat model is applicable in many real-world scenarios.
\subsubsection{Attacker’s Goals}
Typically, attackers aim to ensure the effectiveness of the attack by fooling the model into making targeted mispredictions on poisoned test images while maintaining accuracy on clean datasets. Additionally, robust backdoor attacks should meet the following criteria:
\begin{itemize}
\item[$\bullet$] The backdoor embedded in the model has minimal impact on the victim model's test accuracy on clean samples, while triggered samples should be misclassified into the target class with high probability.
\item[$\bullet$] Triggered samples should appear natural and indistinguishable from the original images to evade manual inspection.
\item[$\bullet$] Attacks should remain effective against existing mainstream defenses.
\end{itemize}
\subsection{Problem Formulation}
This paper investigates the effectiveness of backdoor attacks on image classification neural networks, focusing on typical supervised image classification widely applied in security-sensitive areas such as face recognition and traffic sign recognition.The mapping function for image classification can be described as \( f_{\theta} : \chi \to C \), where \( \chi \) is the input image domain, \( C \) is the output label set, \(  \theta \) represents model parameters that the model can be trained on in the training set, where the original training set can be represented as \( D = \{ (x_i, y_i) \}, x_i \in \chi, y_i \in C \), and \( i = 1, 2, 3, ..., N \). The proposed attack method in this paper is implemented by contaminating the dataset, where the poisoned dataset is denoted as \( D_p \) and the clean dataset as \( D_c \), and finally, the training dataset becomes \( D_t = D_p + D_c \). The model trained on \( D_t \) eventually obtains the model $\hat{f}_{\theta}$, where the poisoning rate is \( p = \frac{D_p}{D} \). The victim model behaves normally in benign datasets but exhibits targeted prediction errors in poisoned datasets that is:
\begin{equation}
f_{\theta}(x_{i}) = y_{i}, \hat{f}_{\theta}(T(x{i})) = y_{t}
\end{equation}
Where $x_i$ represents clean images, $T(x_i)$ represents poisoned images, $y_i$ denotes labels corresponding to clean images, and $y_t$ is the target label set by the attacker. This paper focuses on studying $T(x)$, trigger generation and injection patterns of triggers.
\subsection{Our Attack Overview}

The Singular Value Decomposition (SVD) of an image is a method to decompose the image matrix into singular values and corresponding singular vectors. This process helps us understand the structure of the image and is commonly used for tasks such as image compression, denoising, and feature extraction. Specifically, for an image matrix $A$ of size $m \times n$, where $m$ is the height and $n$ is the width of the image, performing SVD on the image matrix yields the following decomposition:
\begin{equation}
A_{m \times n}=U_{m \times m} \Sigma_{m \times n} V_{n \times n}^{\mathrm{T}}\label{e2}
\end{equation}
Here, $U$ is an $m \times m$ unitary matrix (left singular vector matrix), $\Sigma$ is an $m \times n$ diagonal matrix, and $V$ is an $n \times n$ unitary matrix (right singular vector matrix). The elements on the diagonal of the matrix $\Sigma$ are called the singular values of the image. Larger singular values correspond to capturing the most significant structures and variations in the data, often containing the main features and important signals in the image. Smaller singular values correspond to capturing smaller structural information, but they are less important.

\begin{figure*}[!t]
\centering
\subfloat[On the RGB channels]{
		\includegraphics[width=5.5in,scale=0.5]{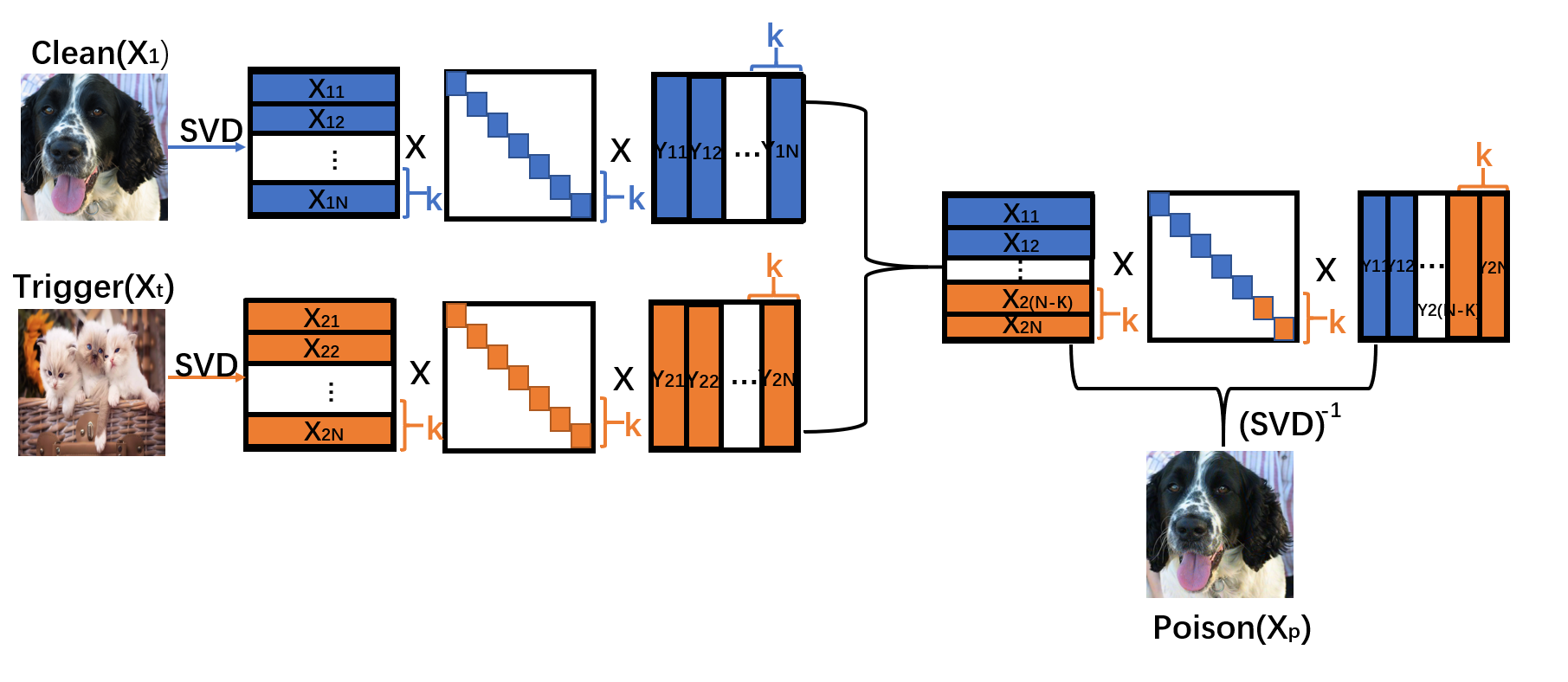}}\\
\subfloat[On the YUV channels]{
		\includegraphics[width=5.5in,scale=0.5]{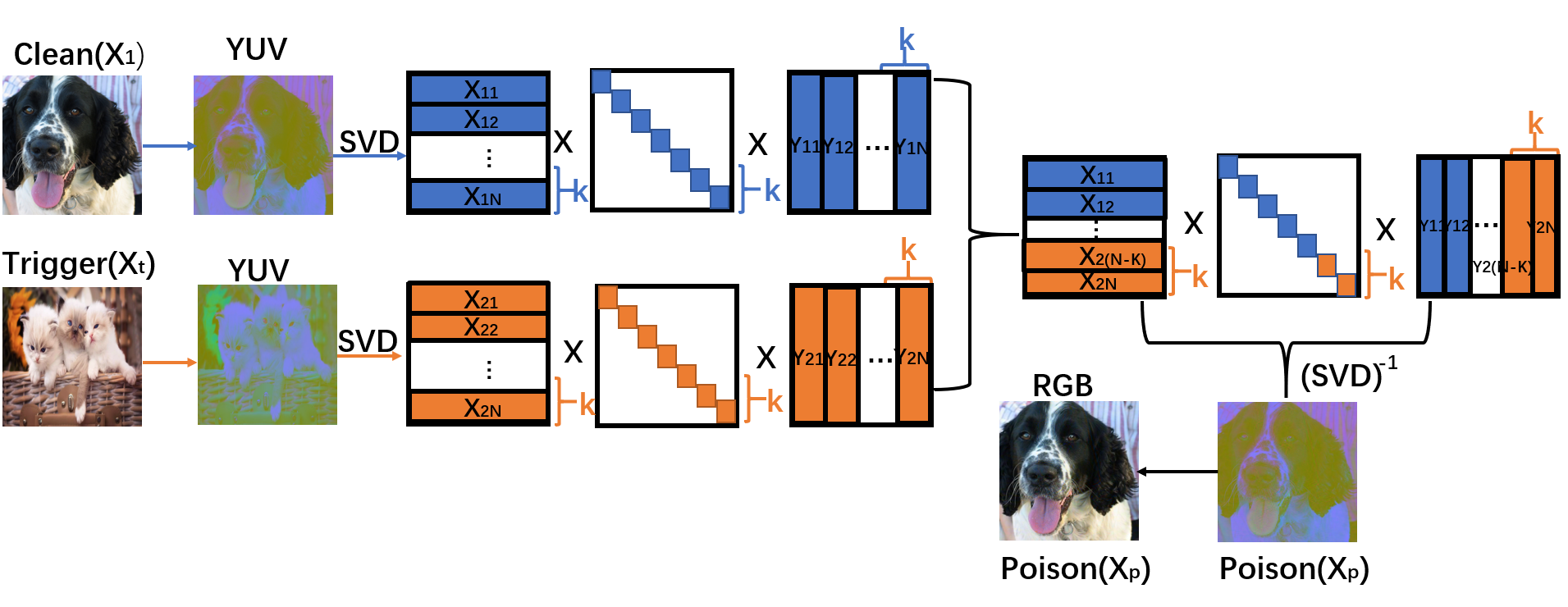}}
\caption{The overview framework of SVD-based backdoor attack (DEBA). Injecting toxic features via SVD on the RGB channels is illustrated in (a). Injecting toxic features via SVD on the YUV channels is illustrated in (b).}
\label{fig.2}
\end{figure*}

The two attack architectures of DEBA shown in Fig \ref{fig.2}, where the first one operates in the RGB channel and the second one operates in the more covert UV channel. Given a clean image $x_{i}\in D$ and a specific trigger image $x_{t}$, we obtain their image matrix decompositions through Equation\ref{e2} as follows:
\begin{equation}
\begin{cases}  & A_{xi}=U_{xi} \Sigma_{xi} V_{xi}^{\mathrm{T}}  \\  & A_{xt}=U_{xt} \Sigma_{xt} V_{xt}^{\mathrm{T}}\end{cases}
\end{equation}

The term $A_{xi}$ represents the result of Singular Value Decomposition (SVD) applied to clean images, while $A_{xt}$ represents the result of Singular Value Decomposition (SVD) applied to triggered images.

Due to the decreasing distribution of singular values in the singular matrix, from the upper left to the lower right, and with most singular values concentrated in the first few positions along the diagonal, while subsequent singular values are very close to zero, the corresponding singular vectors capture less important details of the image. It becomes possible to replace the last $k$ singular values and corresponding singular vectors of the clean image with those of the trigger image, forming a poisoned image. The generated poisoned image retains most of the main features of the clean image and only contains a small amount of secondary features from the trigger image as the trigger, as shown below:
\begin{equation}
\begin{cases}  &  U_p = \begin{bmatrix}U_1, U_2, \cdots , U_{m-k} , U_{m-k+1}^t , \cdots , U_m^t \end{bmatrix} \\  & \Sigma_p = \text{diag}(\sigma_1, \sigma_2, \ldots, \sigma_{m-k}, \sigma_{m-k+1}^t,  \ldots, \sigma_m^t)\\ & V_p^T = \begin{bmatrix}V_1, V_2, \cdots , V_{n-k} , V_{n-k+1}^t , \cdots , V_n^t\end{bmatrix}\end{cases}\label{e4}
\end{equation}
where $U\in U_{xi},U^t\in U_{xt}$ ,$\sigma \in \Sigma_{xi},\sigma^t \in \Sigma_{xt}$,$V\in V_{xi},V^t\in V^T_{xt}$. After Equation\ref{e4} , we obtain the left singular vector matrix $U_P$, the diagonal matrix  $\Sigma_p$, and the right singular vector matrix $V_p$ of the poisoned image. Then, after the inverse singular value decomposition, we finally obtain the poisoned image as shown below:
\begin{equation}
    U_p \Sigma_p V_p^T=x_p
\end{equation}

Where $x_p$ represents the generated poisoned image, retaining most of the main features of the clean image, merely losing some noise and unimportant details, and then incorporating some limited minor features from the trigger image into the clean image. Since what is integrated are some minor features or noise from the trigger image, the poisoned image is visually indistinguishable from the clean image, achieving a truly covert backdoor attack.

\begin{table}[h]
\centering
\caption{Dataset information}
\label{tab1}
\begin{tabular}{llll}
\hline
Datasets      & Train/Test   & Classes & Image Size \\ \hline
CIFAR-10      & 50000/10000  & 10      & 32*32*3    \\
GTSRB         & 39209/12603  & 43      & 64*64*3    \\
Tiny-ImageNet & 100000/10000 & 200     & 64*64*3    \\
ImageNet      & 12480/2496   & 10      & 224*224*3  \\ \hline
\end{tabular}
\end{table}

\section{EXPERIMENTS}
\subsection{Experiment Settings}
\subsubsection{Dataset}
Our targeted attacks mainly focus on neural network classification tasks. To demonstrate the performance of DEBA across different tasks, we conducted experiments on four commonly used classification datasets. These datasets include CIFAR-10\cite{alex2009learning}, GTSRB\cite{stallkamp2012man}, Tiny-ImageNet\cite{abai2019densenet}, and ImageNet\cite{deng2009imagenet}. Among them, GTSRB is a dataset for traffic sign recognition, while the others are object classification datasets. Due to the large size of ImageNet, we only utilized a subset of it for training and testing. Detailed information about these datasets is presented in Table\ref{tab1}.
\subsubsection{Attacked Models}
We conducted experiments on two mainstream classification models, namely ResNet18\cite{he2016deep} and DenseNet121\cite{abai2019densenet}. ResNet18 is a classic classification network, while DenseNet121 is more complex in structure and has more parameters compared to ResNet18.
\subsubsection{baseline Backdoor Attacks}
We compared our method DEBA with existing backdoor methods, including BadNets\cite{gu2019badnets}, Blend\cite{chen2017targeted}, SIG\cite{barni2019new}, and FIBA\cite{feng2022fiba}. BadNets and Blend are classical visible backdoor attacks. SIG is an invisible backdoor attack in the spatial domain, while FIBA is a recent backdoor attack proposed in the frequency domain.
\subsubsection{Evaluation Metrics}
We evaluate DEBA compared to other attack methods from two perspectives: attack performance and attack stealthiness. We use Attack Success Rate (ASR) and Clean Data Accuracy (CDA) to assess the model's attack performance. ASR is defined as the proportion of poisoned examples misclassified as the target label among all poisoned examples used for testing, while CDA indicates the accuracy of the victim model on the clean testing dataset. The formula for computing ASR and CDA are as follows:
\begin{equation}
    \begin{array}{l}A S R=\frac{\sum_{i=1}^{N_{p}} \mathbb{I}\left(f\left(\mathbf{x}_{i}\right)=y_{t}\right)}{N_{p}} \times 100 \% \\C D A=\frac{\sum_{i=1}^{N_{c}} \mathbb{I}\left(f\left(\mathbf{x}_{i}\right)=y_{i}\right)}{N_{c}} \times 100 \% \\\end{array}
\end{equation}
where $N_p$ represents the total number of poisoned samples, $N_c$ represents the total number of clean samples, $f(\textbf{x}_i)$ represents the prediction result of model $f$ on input sample $\textbf{x}_i$, $y_t$ represents the target label set by the attacker, $y_i$ represents the true label of clean sample $\textbf{x}_i$, and $\mathbb{I}(\cdot)$ is the indicator function, which equals 1 when the condition inside the parentheses is true, and 0 otherwise.

An important factor in the stealthiness of covert backdoor attacks is the similarity between the original and poisoned images. Higher similarity leads to the inability of the naked eye and existing defenses to distinguish between poisoned and clean images. For assessing attack stealthiness, we employ the following similarity metrics: Peak Signal-to-Noise Ratio (PSNR)\cite{hore2010image}, Structural Similarity Index (SSIM)\cite{ye2017robust}, and Learned Perceptual Image Patch Similarity (LPIPS)\cite{zhang2018unreasonable}. There is a certain correlation among these three metrics. Generally, an increase in PSNR and SSIM indicates an improvement in image similarity, while a decrease in LPIPS suggests the same.

PSNR (Peak Signal-to-Noise Ratio) is a commonly used metric for assessing image quality. It evaluates the degree of distortion in an image by comparing the ratio of signal to noise between the original image and the poisoned image. PSNR is calculated based on the pixel values of the image and is typically expressed in decibels (dB), with higher values indicating better image quality. The formula for computing PSNR is as follows:
\begin{equation}
P S N R=10 \cdot \log _{10}\left(\frac{\mathrm{MAX} ^{2}}{\mathrm{MSE}}\right)
\end{equation}
where MAX represents the maximum possible pixel value of the image, and MSE (Mean Squared Error) is the average squared difference between the original and processed images at the pixel level.

SSIM (Structural Similarity Index) is a metric used to measure the similarity between two images. It takes into account not only the brightness and contrast of the images but also their structural information.The SSIM value range from 0 to 1, with a higher value
indicating a greater similarity between the images. It is defined as follows:
\begin{equation}
    S S I M(x, y)=\frac{\left(2 \mu_{x} \mu_{y}+c_{1}\right)\left(2 \sigma_{x y}+c_{2}\right)}{\left(\mu_{x}^{2}+\mu_{y}^{2}+c_{1}\right)\left(\sigma_{x}^{2}+\sigma_{y}^{2}+c_{2}\right)}
\end{equation}
where $x$ and $y$ represent the two images being compared. $\mu_x$ and $\mu_y$ are their respective means, $\sigma_x^2$ and $\sigma_y^2$ are their variances, and $\sigma_{xy}$ is their covariance. $c_1$ and $c_2$ are constants used for stability.

LPIPS is a perceptual similarity metric based on deep learning. This method effectively measures the similarity between clean and poisoned images from a human perspective. LPIPS takes into account human perception and is trained to predict human judgments of image similarity. It computes the distance between feature representations of images in deep neural networks. Higher LPIPS values indicate greater dissimilarity between images, while lower values indicate greater similarity. In this paper, we adopt LPIPS based on features learned by pre-trained AlexNet, following the same setup as in the original paper.

\subsubsection{Implementation Details}
In our experiments, we randomly select an image of a cat as the initial trigger image. During the attack phase, we conduct numerous experiments to demonstrate the optimal value of $k$. Since the sizes of images in different datasets vary, the included minor features differ as well. Therefore, it is impossible to fix the value of k and apply it to each dataset uniformly. Through extensive experimentation, our method DEBA (RGB) achieves the optimal $k$ values of 16 for CIFAR-10, 33 for Tiny-ImageNet and GTSRB, and 163 for ImageNet. Our second method DEBA (UV), embedding features in the more invisible UV channels, requires slightly larger $k$ values than the first method. For DEBA (UV), the optimal $k$ values are 26 for CIFAR-10, 41 for GTSRB, 52 for Tiny-ImageNet, and 207 for ImageNet. During the training phase, we use the SGD optimizer to train the victim model with 200 epochs. The learning rate is set to 0.05 with a decay factor of 0.1 at decay steps of 50, 100, and 150, and a batch size of 256. All attacks are configured as All-to-One attacks with a poisoning probability of $p=0.1$. To simulate scenarios closer to real-world conditions, many models nowadays are initially trained using pre-trained models to fine-tune their own architectures. Our network also employs transfer learning from pre-trained models. Therefore, our model may not achieve the highest accuracy on clean datasets.

\subsection{Attack Performance Evaluation}
\subsubsection{Attacks Effectiveness}
The relevant results of DEBA attack are summarized in Table \ref{tab2}. where we assess the effectiveness of different backdoor attacks using ASR and CDA metrics. This indicates that our proposed DEBA is advantageous over these classical attack algorithms on most datasets and models. However, in certain cases, such as experiments on ImageNet and Tiny-ImageNet with specific networks, the ASR of DEBA is slightly lower than BadNets, almost negligible. Given that the trigger meticulously crafted by DEBA is imperceptible in the spatial domain, such results are acceptable. Additionally, compared to the clean baseline, DEBA only incurs a CAD loss below $1$\%. These results suggest that DEBA achieves the desired attack effect.

\begin{table*}[]
\centering
\caption{Comparison of ASR and CDA of DEBA with other
attacks on different models and datasets.}
\label{tab2}
\begin{tabular}{cc|cccccccc}
\hline
\multirow{2}{*}{Models}                           & {Dataset $\rightarrow$}   & \multicolumn{2}{c}{CIFAR-10} & \multicolumn{2}{c}{GTSRB} & \multicolumn{2}{c}{Tiny-ImageNet} & \multicolumn{2}{c}{ImageNet} \\
                                                  & {Attacks $\downarrow$}   & CDA(\%)       & ASR(\%)      & CDA(\%)     & ASR(\%)     & CDA(\%)         & ASR(\%)         & CDA(\%)       & ASR(\%)      \\ \hline
\multicolumn{1}{c|}{\multirow{7}{*}{ResNet18}}    & Clean     & 86.13         & ——           & 96.19       & ——          & 50.13           & ——              & 92.48         & ——           \\
\multicolumn{1}{c|}{}                             & BadNets   & 85.20         & 99.99        & 95.94       & 100.00      & 49.05           & 99.98           & 91.49         & 99.97        \\
\multicolumn{1}{c|}{}                             & Blend     & 85.59         & 99.89        & 96.12       & 99.98       & 49.30           & 99.95           & 91.46         & 99.82        \\
\multicolumn{1}{c|}{}                             & SIG       & 85.63         & 99.92        & 96.18       & 99.91       & 49.38           & 99.93           & 92.33         & 99.92        \\
\multicolumn{1}{c|}{}                             & FIBA      & 84.62         & 98.23        & 96.05       & 99.75       & 47.42           & 95.75           & 90.39         & 96.67        \\
\multicolumn{1}{c|}{}                             & DEBA(RGB) & 85.39         & 99.99        & 96.10       & 100.00      & 48.99           & 99.95           & 92.48         & 99.77        \\
\multicolumn{1}{c|}{}                             & DEBA(UV)  & 85.40         & 99.97        & 95.93       & 99.98       & 49.19           & 99.93           & 91.54         & 99.94        \\ \hline
\multicolumn{1}{c|}{\multirow{7}{*}{DenseNet121}} & Clean     & 88.10         & ——           & 97.02       & ——          & 57.72           & ——              & 92.50         & ——           \\
\multicolumn{1}{c|}{}                             & BadNets   & 87.48         & 100.00       & 96.87       & 100.00      & 55.92           & 99.97           & 92.20         & 99.94        \\
\multicolumn{1}{c|}{}                             & Blend     & 87.06         & 99.85        & 96.61       & 99.98       & 55.73           & 99.94           & 92.30         & 98.79        \\
\multicolumn{1}{c|}{}                             & SIG       & 86.97         & 99.89        & 96.39       & 100.00      & 56.21           & 99.91           & 92.17         & 99.90        \\
\multicolumn{1}{c|}{}                             & FIBA      & 85.64         & 99.63        & 96.36       & 99.84       & 55.65           & 97.98           & 90.57         & 92.78        \\
\multicolumn{1}{c|}{}                             & DEBA(RGB) & 87.68         & 100.00       & 96.57       & 100.00      & 55.94           & 99.50           & 92.61         & 99.79        \\
\multicolumn{1}{c|}{}                             & DEBA(UV)  & 87.32         & 99.98        & 96.53       & 99.98       & 56.60           & 99.91           & 91.74         & 99.92        \\ \hline
\end{tabular}
\end{table*}

\begin{table*}
\centering
\caption{Stealthiness of different attacks in the spatial domain.}
\label{tab3}
\begin{tabular}{c|cccccccccccc}
\hline
\multirow{2}{*}{\begin{tabular}[c]{@{}c@{}}{Dataset $\rightarrow$}\\ {Attacks $\downarrow$}\end{tabular}} & \multicolumn{3}{c}{CIFAR-10} & \multicolumn{3}{c}{GTSRB} & \multicolumn{3}{c}{Tiny-ImageNet} & \multicolumn{3}{c}{ImageNet} \\
                                                                                & PSNR    & SSIM    & LPIPS    & PSNR   & SSIM   & LPIPS   & PNSR      & SSIM      & LPIPS     & PSNR    & SSIM    & LPIPS    \\ \hline
BadNets                                                                         & 25.21   & 0.969   & 0.0069   & 27.87  & 0.978  & 0.0078  & 24.42     & 0.971     & 0.0871    & 26.02   & 0.984   & 0.0326   \\
Blend                                                                           & 21.63   & 0.876   & 0.0164   & 20.90  & 0.854  & 0.0575  & 18.89     & 0.732     & 0.1227    & 20.22   & 0.771   & 0.3118   \\
SIG                                                                             & 24.24   & 0.890   & 0.0391   & 24.01  & 0.908  & 0.0097  & 24.04     & 0.831     & 0.0226    & 25.15   & 0.921   & 0.1445   \\
FIBA                                                                            & 31.14   & 0.976   & 0.0025   & 28.32  & 0.987  & 0.0024  & 29.48     & 0.990     & 0.0059    & 29.03   & 0.976   & 0.0248   \\
DEBA(RGB)                                                                       & 35.16   & 0.988   & 0.0003   & 37.99  & 0.992  & 0.0001  & 32.66     & 0.963     & 0.0018    & 34.40   & 0.913   & 0.0248   \\
DEBA(UV)                                                                        & 36.98   & 0.993   & 0.0001   & 44.78  & 0.998  & 4.9779  & 40.10     & 0.994     & 0.0005    & 36.55   & 0.978   & 0.0126   \\ \hline
\end{tabular}
\end{table*}

\subsubsection{Attack Stealthiness}
\begin{figure}[!t]
\centering
\includegraphics[width=3.5in]{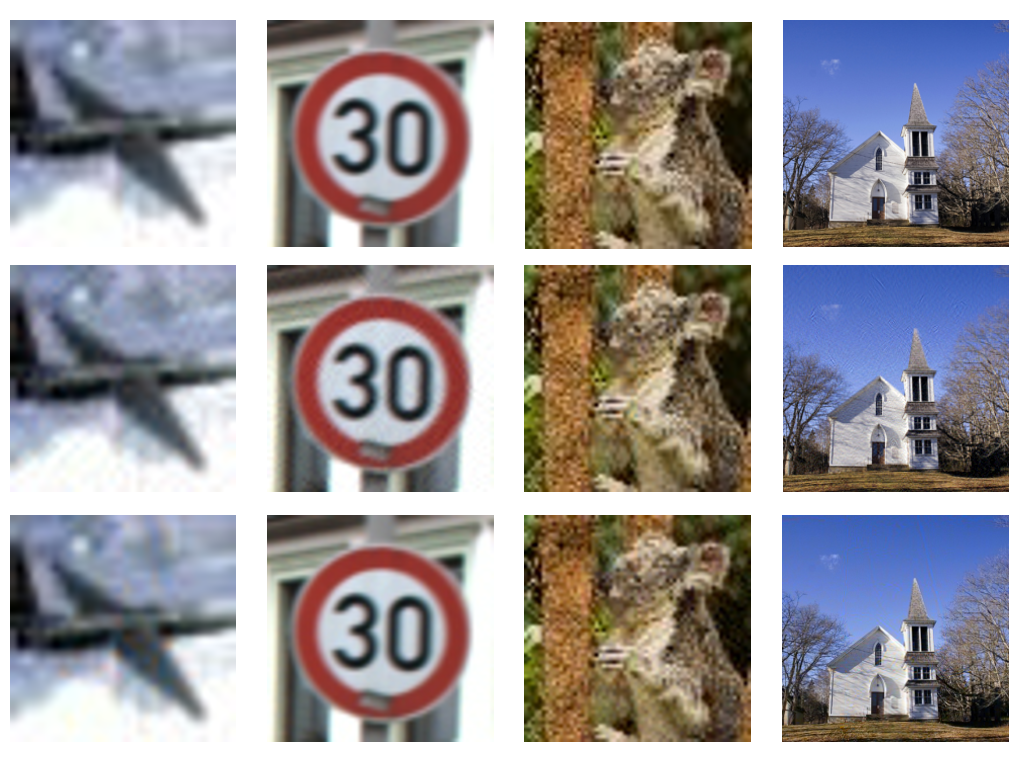}
\caption{
The visualization of stealthiness on different datasets. The first row consists of clean original images, the second row shows toxic images generated by DEBA(RGB), and the third row displays toxic images generated by DEBA(UV). Images from left to right columns are respectively from the datasets: CIFAR-10, GTSRB, Tiny-ImageNet, and ImageNet.}
\label{Fig.3}
\end{figure}

Next, we evaluate the stealthiness of DEBA. As shown in Fig \ref{Fig.3}, we observe strong stealthiness of DEBA across different datasets, where it becomes almost imperceptible to distinguish between clean images and toxic images generated by DEBA. Additionally, as depicted in Fig \ref{Fig.1}, our method demonstrates superior stealthiness compared to other attack algorithms. The toxic images generated by our method appear more natural and visually indistinguishable from clean images compared to other attack methods. In Table \ref{tab3}, we compare our method with others using three evaluation metrics: PSNR, SSIM, and LPIPS. It is evident from the table that both variants of DEBA outperform any other attack method in terms of these evaluation metrics across different datasets, with DEBA(UV) particularly excelling in each metric. In summary, DEBA achieves the best stealth results through comprehensive visual perception and various metrics.

\subsubsection{Poisoning Rate}
\begin{figure}[!t]
\centering
\includegraphics[width=3.5in]{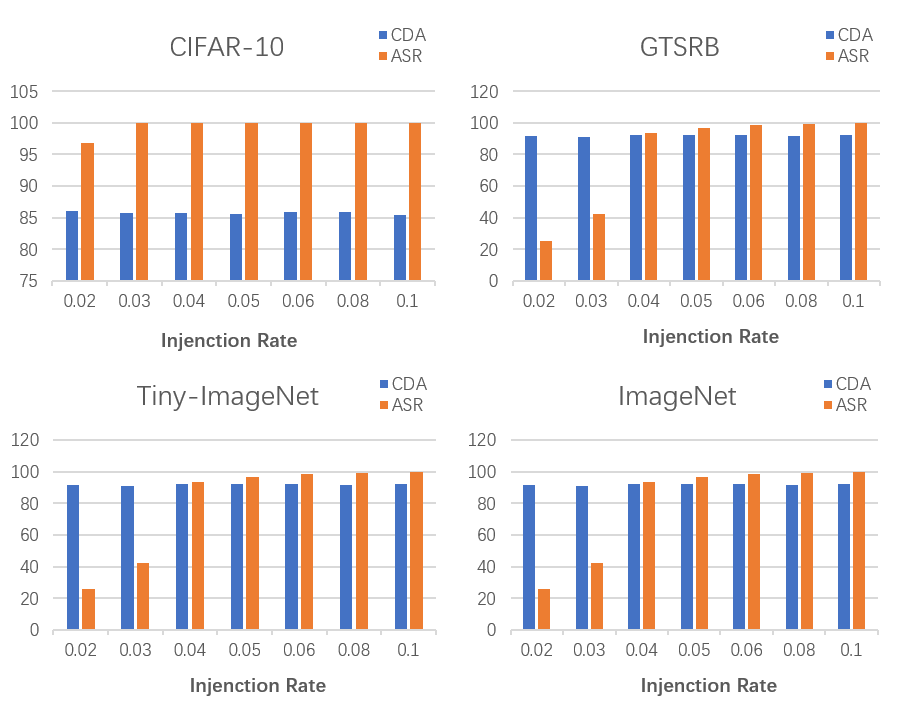}
\caption{
The impact of DEBA on attack performance at different injection rates across various datasets.}
\label{Fig.4}
\end{figure}

Next, we evaluated the effectiveness of DEBA under varying injection rates of toxic images in the training data. We applied DEBA(RGB) to different datasets and conducted experiments on ResNet18. As shown in Fig \ref{Fig.4} on CIFAR-10, an injection rate as low as 0.02 already achieves an ASR of over 95\%, while on the other three datasets, ASR reaches close to 100\% at an injection rate of 0.04, maintaining stable CAD. This experiment validates the effectiveness of our DEBA, achieving high ASR at low injection rates without sacrificing the accuracy of the poisoned model on clean datasets.

\subsubsection{Trigger Magnitude}
The magnitude of the trigger also affects the attack performance of DEBA. In this study, the trigger magnitude is primarily determined by the value of $k$, where a larger $k$ corresponds to a larger trigger magnitude, indicating that more features of the clean image are replaced by those of the trigger image. We mainly investigated the effect of different values of $k$ on the performance of DEBA (RGB) on the CIFAR-10 dataset. As shown in Table \ref{tab4}, we conducted experiments by changing the value of $k$ from 12 to 17. When $k$ was set to 12, the ASR reached over 99\% on both ResNet18 and DenseNet121 models, approaching 100\%, without significant loss in CAD. When k was set to 17, the attack performance peaked at 100\% on both models, with minimal loss in CAD.

\begin{table}[]
\centering
\caption{The impact of different values of $k$ on the attack performance on the CIFAR-10 dataset.}
\label{tab4}
\begin{tabular}{cccll}
\hline
\multirow{2}{*}{\begin{tabular}[c]{@{}c@{}}{Model $\rightarrow$}\\ {Value of k$\downarrow$}\end{tabular}} & \multicolumn{2}{c}{ResNet18} & \multicolumn{2}{l}{DenseNet121} \\
                                                                            & CDA          & ASR           & CDA            & ASR            \\ \hline
\multicolumn{1}{c|}{0}                                                      & 86.13        & ——            & 88.10          & ——             \\
\multicolumn{1}{c|}{12}                                                     & 85.59        & 99.88         & 87.12          &99.92                \\
\multicolumn{1}{c|}{13}                                                     & 85.42        & 99.96         & 86.32          & 99.99               \\
\multicolumn{1}{c|}{14}                                                     & 85.56        & 99.98         & 87.18          & 99.99          \\
\multicolumn{1}{c|}{15}                                                     & 85.28        & 99.98         & 87.35          & 99.99          \\
\multicolumn{1}{c|}{16}                                                     & 85.39        & 99.99         & 87.68          & 100.00         \\
\multicolumn{1}{c|}{17}                                                     & 85.20        & 100.00        & 87.18          & 100.00         \\ \hline
\end{tabular}
\end{table}
\section{CONCLUSION}
In this paper, we demonstrate that most backdoor attacks are visible in the spatial domain. To fully circumvent existing defenses while maintaining spatial invisibility, we propose an invisible backdoor attack in the spatial domain, called DEBA. DEBA leverages the mathematical properties of Singular Value Decomposition (SVD) to embed the minor features of trigger images as backdoor information, using the decomposed singular values and vectors. Additionally, to enhance the efficiency and invisibility of DEBA, we conduct feature embedding in the UV channels. Through extensive experimental evaluations, we validate the superior performance of DEBA in terms of attack success rate and attack invisibility. We believe that our work contributes to the development of more robust and secure image classification neural networks.


%



\ifCLASSOPTIONcaptionsoff
  \newpage
\fi



%


\bibliographystyle{IEEEtran}
\bibliography{Mybbl}

%




\end{document}